\RequirePackage{fix-cm}
\documentclass[smallextended]{svjour3}       
\def\makeheadbox{\relax}
\smartqed  
\newcommand{\eq}[1]{\begin{eqnarray}#1\end{eqnarray}}

  \usepackage{graphicx}
  \usepackage{amssymb, amsmath}
  \usepackage{graphicx}
  \usepackage{enumitem}
  \usepackage{caption}
  \usepackage{mathtools}
  \usepackage{color}
  \usepackage{hyperref}
  \usepackage{bm}
  \usepackage[numbers]{natbib}

\begin{document}

\title{Collisions near Kerr black holes: 
lower limit of energy between orbiting and incoming particles}
\titlerunning{Collisions near Kerr black holes: 
lower limit of energy...}

\author{Mieszko Rutkowski
}


\institute{M. Rutkowski \at
              Marian Smoluchowski Institute of Physics, Jagiellonian University\\
              \L ojasiewicza 11, 30-348 Cracow, Poland\\
              \email{\href{mailto:mieszko.rutkowski@student.uj.edu.pl}{mieszko.rutkowski@student.uj.edu.pl}}           
}

\date{December 2016}

\maketitle

\begin{abstract}
In our paper we investigate the lower limit of collisional energy of test particles near the Kerr black hole. In particular we examine the minimal Lorentz factor between the freely falling particles
and the particles orbiting around a black hole. We consider collisions on the innermost stable circular orbit (\textit{ISCO}) and examine near--extreme case,
where collisions take place near an event horizon. By fine--tuning the particles' angular momentum, the Lorentz factor of the collision can always be minimized to a value dependent on the
black hole's spin. We identified that this minimal value is always less than $\frac{2\sqrt{2}-1}{\sqrt{3}}$ and more than $\frac{\sqrt{12}-1}{\sqrt{6}}$ (the limits are
the values for an extreme Kerr and Schwarzschild, respectively).
It implies that this kind of collisions of compact objects are expected to be highly energetic near supermassive black holes.
In addition, we show that an interaction between black hole's and particle's spins has an influence
on minimal Lorentz factor. This contribution is nonnegligible for near--extreme black holes.
We also discuss the relation between our results and sci--fi movie \textit{Interstellar}.
\end{abstract}

\section{Introduction}

Collisions in the ergosphere were in the centre of interest, because
they are able to extract rotational energy from the black hole through a collisional Penrose process. As described by \cite{Ban09},
collisional energy of particles near the
extreme-Kerr black hole can be arbitrarily high, although, this does not imply that such collisions can extract arbitrarily high amounts of energy from a black hole. \cite{Be12} showed
that the energy of the particles leaving the ergosphere after a collision and measured by distant observers, is in fact not significantly higher
than the primary energy of infalling particles.
Despite their doubtful application as cosmic accelerators, collisions around black holes
are interesting in light of compact object mergers. For example, collisions of neutron stars could
possibly lead to the birth of a black hole (for numerical simulations see e.g.\ \cite{Go12}). 

General formulas for the CM energy of collisions (energy in the centre of mass frame) were shown by \cite{Ha14},
collisions on ISCO were investigated by \cite{Ha11}, and collisions of spinning particles by \cite{Gu16}. In these papers the authors showed
that an upper bound of CM collisional energy diverges to infinity for the extreme-Kerr black holes,
and \cite{Ha14} concluded that this would result in the formation of a black hole in case of a NS-NS collision.

Since upper limits of collisional energy have been investigated in many ways, no discussion on lower limit of collisional energy has been presented so far.
Our paper fills this gap.
Solving this issue leads to a full energy domain of collisions: that would have an application
in studies on compact objects collisions, for example as an initial value in numerical simulations.
We derive the formulas that minimize collisional energy in case of one particle orbiting around a black hole, and another particle infalling freely from infinity.
Furthermore, we present how the situation changes when an infalling particle has a nonzero spin - it has been investigated recently by \cite{Gu16} in context of upper energy limits.
We show that effects of interaction between particle's and black hole's spins affects minimal collisional energy in near--extreme Kerr background only.

Resolving the issue of minimal energy is interesting also in light of sci--fi movies, like \textit{Interstellar}. Treating a spaceship as a incoming particle and a planet as an orbiting particle,
we can calculate the minimal speed between them. Therefore we can predict wheather planets near black holes are ``available'' for a spaceship to land on them.

The second section of our paper introduces notation and describes the motion in the Kerr space--time domain. The third section presents derivations of formulas for the minimal
Lorentz factor of collisions and discusses the special case of the collisions on ISCO. The fourth section introduces Mathisson--Papapetrou equations and contains a discussion of impact of particle's spin
on a situation.
The last section gives a brief summary of an article.

\section{Geodesics in Kerr metric}\label{sec2}
Kerr space--time can be described by the Boyer--Linquist coordinates \cite{Bo67} with a line element given by:
\eq{
g=-\frac{\Delta-a^2\sin^2{\theta}}{\Sigma}\textrm{dt}^2-\frac{4aMr}{\Sigma}\sin^2{\theta}\textrm{dt}\textrm{d}\phi +\nonumber \\
+\frac{(r^2+a^2)^2-\Delta a^2\sin^2{\theta}}{\Sigma}\sin^2{\theta}\textrm{d}\phi^2+\frac{\Sigma}{\Delta}\textrm{dr}^2+\Sigma \textrm{d}\theta^2\, ,
}
where $\Sigma=r^2+a^2\cos^2{\theta}$ and $\Delta=r^2+a^2-2Mr$, where $M$ and $a$ represent mass and spin of a black hole ($M\geq a$), respectively.
Range of variables is given by: $t\in(-\infty,\infty), r\in(M+\sqrt{M^2-a^2},\infty),\, \theta\in(0,\pi),\, \phi\in [0,2\pi).$
The motion is considered in the equatorial plane only. Time--translational
and rotational Killing fields are as follows, $\xi$=$\partial_t$ and $\eta=\partial_\phi$. Quantities conserved along geodesics associated with Killing fields are specified as below,
\begin{flalign}
&e=-v_\mu \xi^\mu\, ,\label{c1}\\
&l=v_\mu \eta^\mu\, ,\label{c2}
\end{flalign}
where $v^\mu$ indicates 4-velocity of a particle. These constants represent energy at infinity per unit mass and angular momentum parallel to the z-axis per unit mass, respectively.
Using equations \eqref{c1} and \eqref{c2}, we obtain:
\begin{flalign}
&v^t=-\frac{g_{\phi\phi} e+g_{t\phi}l}{\bar g}\, ,\label{ut}\\
&v^\phi=\frac{g_{t\phi} e+g_{tt}l}{\bar g}\, ,\label{ufi}
\end{flalign}
where $\bar g=g_{tt}g_{\phi\phi}-(g_{t\phi})^2=-\Delta$. We consider motion outside the event horizon, so $\bar g<0$.
Another conserved quantity comes from the velocity normalization condition:
\eq{
v^\mu v_\mu=-1\, .\label{c3}
}
Combining equations \eqref{c1}, \eqref{c2} and \eqref{c3} together results in an equality:
\begin{align}
\frac{\dot r^2}{2}-\frac{M}{r}+\frac{a^2(1-e^2)+l^2}{2r^2}-\frac{M(ae-l)^2}{r^3}=\frac{e^2-1}{2}\, .
\end{align}
This can be interpreted as a sum of a kinetic energy and an effective potential, therefore the problem reduces to a problem from classical mechanics with a potential:
\eq{
V_{eff}(r)=-\frac{M}{r}+\frac{a^2(1-e^2)+l^2}{2 r^2}-\frac{M (a e-l)^2}{r^3}\, . 
}
We also provide below another formula that will be useful in the next paragraph of this article. Combining equations \eqref{c1}, \eqref{c2} and \eqref{c3}, we find a relation:
\eq{
g_{\phi\phi}e^2+g_{tt}l^2+2g_{t\phi}el=\Delta[1+g_{rr}(v^r)^2]\,, \label{normel}
}

\section{Lower limit of collisional energy}\label{sec3}
\subsection{General case}
Two particles are considered to be moving along geodesics. Quantities $p_o^\mu=m_ov_o^\mu$ and $p_{in}^\mu=m_{in}v_{in}^\mu$ denotes
momenta of the orbiting and the incoming particle, respectively. The orbiting particle is staying on an orbit at radius $r=r_o$ and the incoming particle is infalling
from infinity ($e_{in}\geq 1$).
Energy of an incoming particle measured in a rest-frame of the first particle is given by:
\eq{
E=- g_{\mu\nu}v_o^\mu p_{in}^\nu \,,\label{energy}
}
On the other hand, in a locally flat coordinate system, this energy is equal to:
\eq{
E=m_{in}\gamma,
}
where $\gamma$ is a Lorentz factor of the collision. Therefore, using equations \eqref{ut} and \eqref{ufi}, we obtain (since now, all the metric coefficients are calculated
at $r=r_o$):
\begin{eqnarray}
\gamma=-\frac{1}{g}(g_{tt}g_{t\phi} v_o^te_{in}+g_{tt}g_{t\phi} v_o^t l_{in}-g_{t\phi}^2 v_o^te_{in} +g_{t\phi}g_{tt}v_o^t l_{in}+\nonumber\\
+g_{t\phi}g_{\phi\phi} v_o^\phi e_{in}+ g_{t\phi}^2 v_o^\phi l_{in} -g_{\phi\phi} g_{t\phi} v_o^\phi e_{in}-g_{\phi\phi}g_{tt}v_o^\phi l_{in})+g_{rr} v_o^r v_{in}^r\,.
\end{eqnarray}
Hence, for a particle staying on a circular orbit ($v_o^r=0$):
\eq{
\gamma= v_o^t e_{in}- v_o^\phi l_{in}\label{g0}\,.
}
This remarkably simple formula is useful especially when components $v_o^t$ and $v_o^\phi$ are fixed quantities, like in our case.
However, one needs to remember that $e_{in}$ and $l_{in}$ have to be chosen in such a way
that world lines of colliding particles should have a common point.
In order to minimize $\gamma$, we have to choose appropriate $e_{in}$ and $l_{in}$. The conditions for $\gamma$ to be minimal are found in the following way:
\begin{enumerate}[label=(\roman*)]
\item Let us find the solution of equation \eqref{normel} with respect to $l$. This implies that for $r\neq 2M$ $l_{in}$ takes the form (we choose co--rotating case):
\eq{
l_{in}=-e_{in}\frac{g_{t\phi}}{g_{tt}}-\sqrt{\Delta}\frac{\sqrt{e_{in}^2+g_{tt}+g_{rr}g_{tt}(v_{in}^r)^2}}{g_{tt}}\,, \label{lmin}
}
therefore $\gamma$ \eqref{g0} is given by:
\eq{
\gamma=v_o^t e_{in}+v_o^\phi e_{in}\frac{g_{t\phi}}{g_{tt}}+v_o^\phi \sqrt{\Delta}\frac{\sqrt{e_{in}^2+g_{tt}+g_{rr}g_{tt}(v_{in}^r)^2}}{g_{tt}}\,.\label{game}
}


For $r=2M$ the solution of the system of equations \eqref{c1}, \eqref{c2} and \eqref{c3}
is shown below:
\eq{
l_{in,2M}=-\frac{\left(e_{in}\right)^2 g_{\phi \phi}-\Delta[1+g_{rr} (v_{in}^r)^2]}{2e_{in} g_{t\phi}}\,,\label{l2m}
}
hence
\eq{
\gamma_{2M}=v_o^t e_{in}+v_o^\phi\frac{\left(e_{in}\right)^2 g_{\phi \phi}-\Delta[1+g_{rr} (v_{in}^r)^2]}{2e_{in} g_{t\phi}}\,.\label{game2m}
}

\item \label{i2}
It can be easily checked that \eqref{game} [and \eqref{game2m}] is minimal for $e_{in}=e_o$ and $v_{in}^r=0$ and is a growing function of $e_{in}$ for $e_{in}>e_o$ ($e_o<1$). However, we imposed
one particle to infall from infinity e.g.\ $e_{in}\geq1$, therefore we take $e_{in}=1$.
Condition $v_{in}^r=0$ is equivalent to the condition stating that trajectory of an infalling particle is tangent to the trajectory of an orbiting particle.

\end{enumerate}

Applying condition \ref{i2} to the relation \eqref{normel}, we obtain:
 \begin{flalign}
 g_{\phi\phi}+g_{tt}l_{in}^2+2g_{t\phi}l_{in}=\Delta\,,\\ 
 g_{\phi\phi}e_o+g_{tt}l_{o}^2+2g_{t\phi}e_ol_{o}=\Delta\,. 
 \end{flalign}
Now, together with \eqref{c1} and \eqref{c2}, minimal $\gamma$ takes the form:
\eq{
\gamma_{min}=\frac{1}{g_{tt}}\left[ \sqrt{\left(1+g_{tt}\right)\left(e_o^2+g_{tt}\right)}-e_o\right]\,,\label{gammamin}
}
which is valid for $r\neq2M$. For $r=2M$ minimal $\gamma$ is given by:
\eq{
\gamma_{min,2M}=\frac{1}{2}\left(e_{o}+\frac{1}{e_{o}}\right)\,.\label{gammamin2m}
}
It can be seen that:
\eq{
\lim\limits_{r\to2M^+}{\gamma_{min}}=\lim\limits_{r\to2M^-}{\gamma_{min}}=\gamma_{min,2M}\,.
}
Using formulas \eqref{lmin} and \eqref{l2m} we can also find $l_{in}$ corresponding with $v^r_{in}(r=r_{o})=0$ and $e=1$. For $r\neq 2M$ it is given by:
\eq{
l_{in,min}=-\frac{g_{t\phi}+\sqrt{\Delta\left(1+g_{tt}\right)}}{g_{tt}}\,,
}
and for $r=2M$:
\eq{
l_{in,min,2M}=\frac{\Delta-g_{\phi\phi}}{2g_{t\phi}}\,.
}

\begin{figure}[h!!!]
\includegraphics[width=\columnwidth]{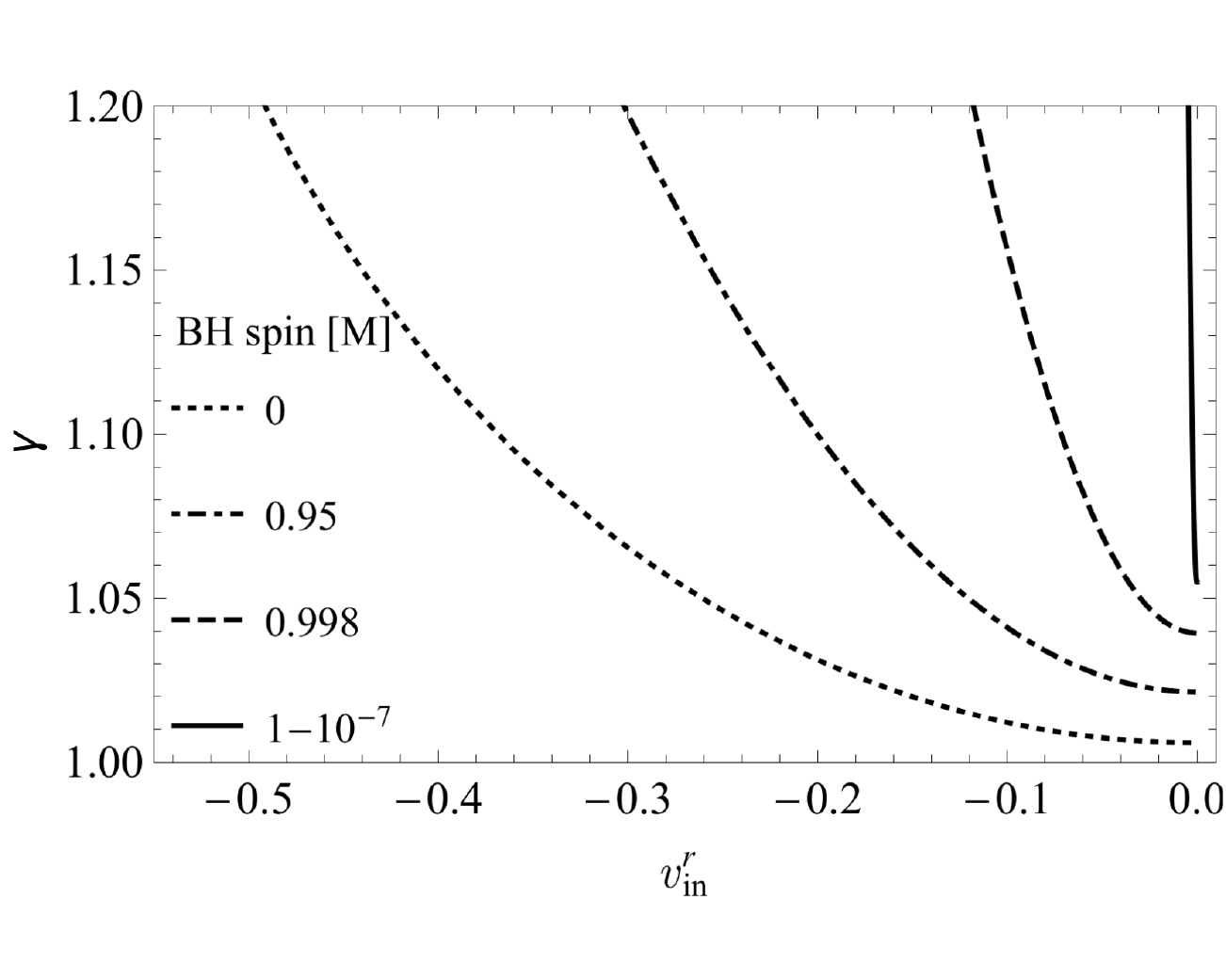}
\caption{Dependence of $\gamma$ on the radial velocity of incoming particle on ISCO. In extreme--Kerr limit, particle's motion has to be fine--tuned in order to have finite value of $\gamma$.}
\label{fig1}
\end{figure}

We note that formula \eqref{gammamin} is valid for Kerr--Newman metric as well. Also, we point out the fact that minimal Lorentz factor calculated above requires fine--tuning
of incoming particle's angular momentum (or $v^r_{in}(r_0)$, equivalently). 
This requirement is presented in Figure \ref{fig1} --- in the extreme--Kerr limit any $v^r_{in}(r_0)\neq 0$ results in infinite energy.

\color{black}
\subsection{Collisions on ISCO}
This section considers the first particle to stay on ISCO. The radius of ISCO may be obtained by deriving the system of the following three equations:
\begin{flalign}
&V_{eff}(r_{isco})=\frac{e^2-1}{2},\\
&V_{eff}'(r_{isco})=0,\\ 
&V_{eff}''(r_{isco})=0.
\end{flalign}
The first two equations are conditions for the circular motion. The third one imposes the potential to have an inflection point at $r_o$. The shape of potential
implies that this corresponds with the marginally stable orbit.
Explicit formula for $r_{isco}$ is to be found in \cite{AB}. Calculating minimal $\gamma$ can be easily done by substituting $r_o=r_{isco}$ and $e_o=e_{isco}$ in equation \eqref{gammamin}
[or \eqref{gammamin2m} for $r=2M$].

Values of $\gamma_{min}$ for characteristic spins of a black hole are to be found in Table \ref{tab}. We note the fact that for $a=M$, the ISCO does not exist \cite{Ha11} and the given value should be understood
as $\lim\limits_{a\to M^-}{\gamma_{min}}$.

\begin{table}
\begin{tabular}{|c|c|c|c|c|}
\hline
$a[M]$&$r_{isco}[M]$&$e_{isco}$&$l_{in,min}[M]$&$\gamma_{min}$\\
\hline
$0$&$6$&$\frac{2\sqrt{2}}{3}$&$3\sqrt{2}$&$\sqrt{2}-\frac{1}{\sqrt{6}}$\\
\hline
$\frac{2\sqrt{2}}{3}$&$2$&$\frac{\sqrt{2}}{\sqrt{3}}$&$\frac{11}{3\sqrt{2}}$&$\frac{5}{2\sqrt{6}}$\\
\hline
$0.998$&$1.237$&$0.679$&$2.145$&$1.039$\\
\hline
$1$&$1$&$\frac{1}{\sqrt{3}}$&$2$&$\frac{2\sqrt{2}-1}{\sqrt{3}}$\\
\hline
\end{tabular}
\caption{Minimal $\gamma$ factors for collisions on ISCO, for some characteristic values of $a$. Value $a=0.998M$ corresponds with a maximal spin of astrophysical black holes predicted by \cite{Th74}.}
\label{tab}
\end{table}

For the near--extreme Kerr black holes, $\gamma_{min}$ can be expressed by an approximate formula, where $\epsilon=\sqrt{M^2-a^2}$:
\eq{
\gamma_{min}\simeq\frac{2 \sqrt2-1}{\sqrt{3}} + \frac{\sqrt[3]{2} \left(2 \sqrt{2}-3\right)}{\sqrt{3}}\epsilon ^{2/3}+\mathcal{O}(\epsilon^{5/3})\,.
}

\section{Spinning particles}\label{sec4}
Motion of particles with spin is described by a set of Mathisson--Papapetrou equations \cite{Mat37,Pap51}:
\begin{flalign}
&\frac{d x^\mu}{d\tau}=v^\mu\label{MP1}\\
&\frac{Dp^\alpha}{d\tau}=-\frac{1}{2}R^\alpha_{\,\,\beta\mu\nu}v^\beta S^{\mu\nu}\label{MP2}\\
&\frac{DS^{\alpha\beta}}{d\tau}=2p^{[\alpha}v^{\beta]}\label{MP3}
\end{flalign}
where $S^{\mu\nu}$ denotes the spin tensor, $v^\mu$ 4-velocity, $p^\mu$ momentum (which is in general not tangent to the particle's world line) and $\frac{D}{d\tau}$ is a covariant derivative
with respect to particle's proper time (e.g. $\frac{D}{d\tau}\equiv v^\mu\nabla_\mu$). These equations are valid for small value of particle's spin parameter $s$ (where $s^2=-S^{\mu\nu}S_{\mu\nu}$).

In case of motion in equatiorial plane in Kerr space--time, the only nonzero components of spin tensor are given by \cite{Hac14}:
\eq{
S^{rt}=-\frac{sp_\phi}{m r},\,\,\,\,\, S^{\phi t}=\frac{s p_r}{m r},\,\,\,\,\, S^{\phi r}=-\frac{Sp_t}{m r}.
}
where $m=-p^\mu p_\mu$. It also means that a particle's spin vector defined as
\eq{
S^\alpha=\frac{1}{2m\sqrt{-g}}\epsilon^{\alpha\beta\mu\nu} p_\beta S_{\mu\nu}
}
is perpendicular to the equatiorial plane.

Method of treating equations \eqref{MP1}, \eqref{MP2}, \eqref{MP3} in equatorial plane is presented
by \cite{Sai98}. The result is given by:
\begin{flalign}
&\Sigma_s\Lambda_s\frac{dt}{d\tau}=a\left(1+\frac{3Ms^2}{r\Sigma_s}\right)[L-(a+s)E]+\frac{r^2+a^2}{\Delta}P_s,\label{S1}\\
&\left(\Sigma_s\Lambda_s\frac{dr}{d\tau}\right)^2=R_s,\label{S2}\\
&\Sigma_s\Lambda_s\frac{d\phi}{d\tau}=\left(1+\frac{3Ms^2}{r\Sigma_s}\right)[L-(a+s)E]+\frac{a}{\Delta}P_s\label{S3},
\end{flalign}
where
\begin{flalign}
&\Sigma_s=r^2(1-\frac{M s^2}{r^3}),\\
&\Lambda_s=1-\frac{3Ms^2r[-(a+s)E+L]^2}{\Sigma_s^3},\\
&R_s=P_s^2-\Delta\left(\frac{\Sigma_s^2}{r^2}+[-(a+s) E+L]^2\right),\\
&P_s=\left((r^2+a^2)+\frac{a s}{r}(r+M)\right)E-\left(a+\frac{Ms}{r}\right)L.
\end{flalign}
$E$ and $L$ are constants of motion given by:
\begin{flalign}
&E=-(p^\mu\xi_\nu-\frac{1}{2}S^{\alpha\beta}\nabla_\beta\xi_\alpha),\\
&L=p^\mu\eta_\nu-\frac{1}{2}S^{\alpha\beta}\nabla_\beta\eta_\alpha.
\end{flalign}

\begin{figure}[h!!!]
\includegraphics[width=\columnwidth]{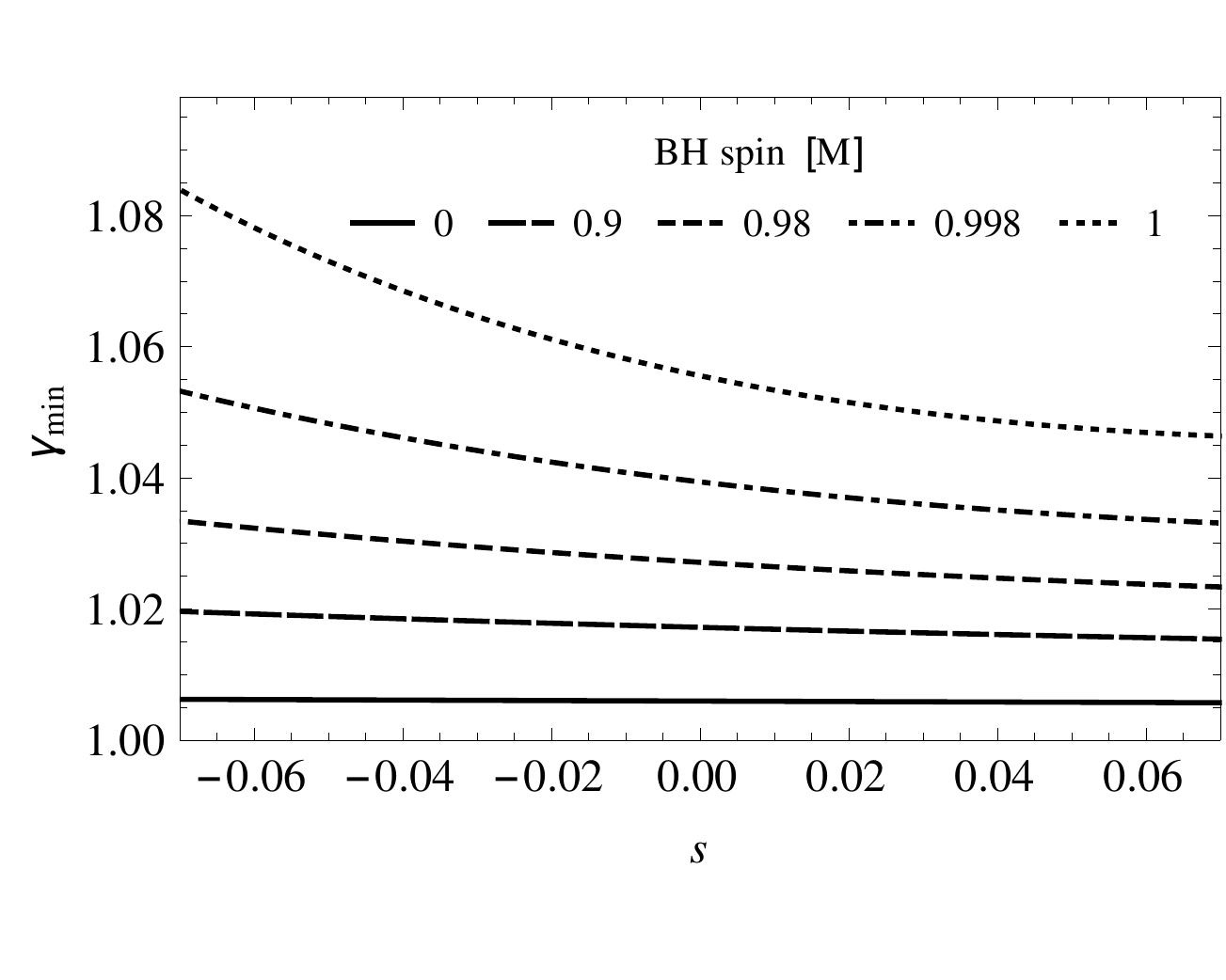}
\caption{Minimal Lorentz factor for collisions on ISCO as a function of incoming particle's spin. The dependence is presented for several values of black hole's spin. Results for $s=0$ correspond with results from section \ref{sec3}.}
\label{fig2}

\end{figure}

These equations may be used to calculate suitable collisional energies.
We present results of numerical calculations of lower bounds of collisional energy.
The calculation was done by solving equations \eqref{S1}-\eqref{S3} with respect to $\frac{d t}{d\tau}$, $\frac{d r}{d\tau}$ and $\frac{d \phi}{d\tau}$ (assuming that incoming particle is at rest at infinity),
and calculating the Lorentz factor using equation:
\eq{
\gamma=-g_{\mu\nu}v^\mu_{o}\frac{d x^\nu}{d\tau}.
}
The orbiting particle was considered to be spinless and to move on ISCO.
Results are presented in Figure \ref{fig2}. It turns out that for parallel oriented spins of black hole and particle the minimal Lorentz factor
is reduced, whereas in an anti--parallel case Lorentz the factor may be significantly higher. However, this effect appears to be significant for
fastly--rotating black holes only (what is partially a consequence of a higher value of ISCO radius for lower spins).
\begin{figure}[h!!!]
\includegraphics[width=.95\columnwidth]{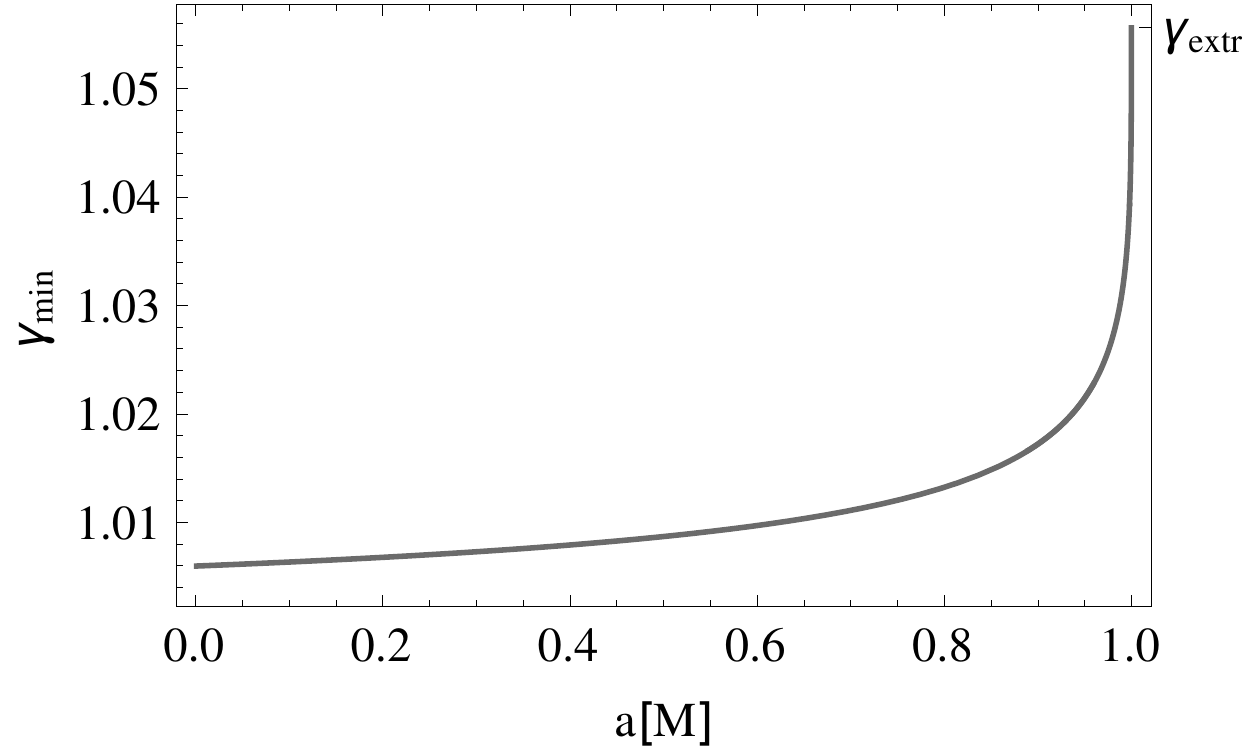}
\caption{Minimal Lorentz factor for collisions on ISCO as a function of black hole's spin. Minimal Lorentz factor for the extreme--Kerr limit is denoted by $\gamma_{extr}=\frac{2\sqrt{2}-1}{\sqrt{3}}$.}
\label{fig3}
\end{figure}

\begin{figure}[h!!!]
\includegraphics[width=.87\columnwidth]{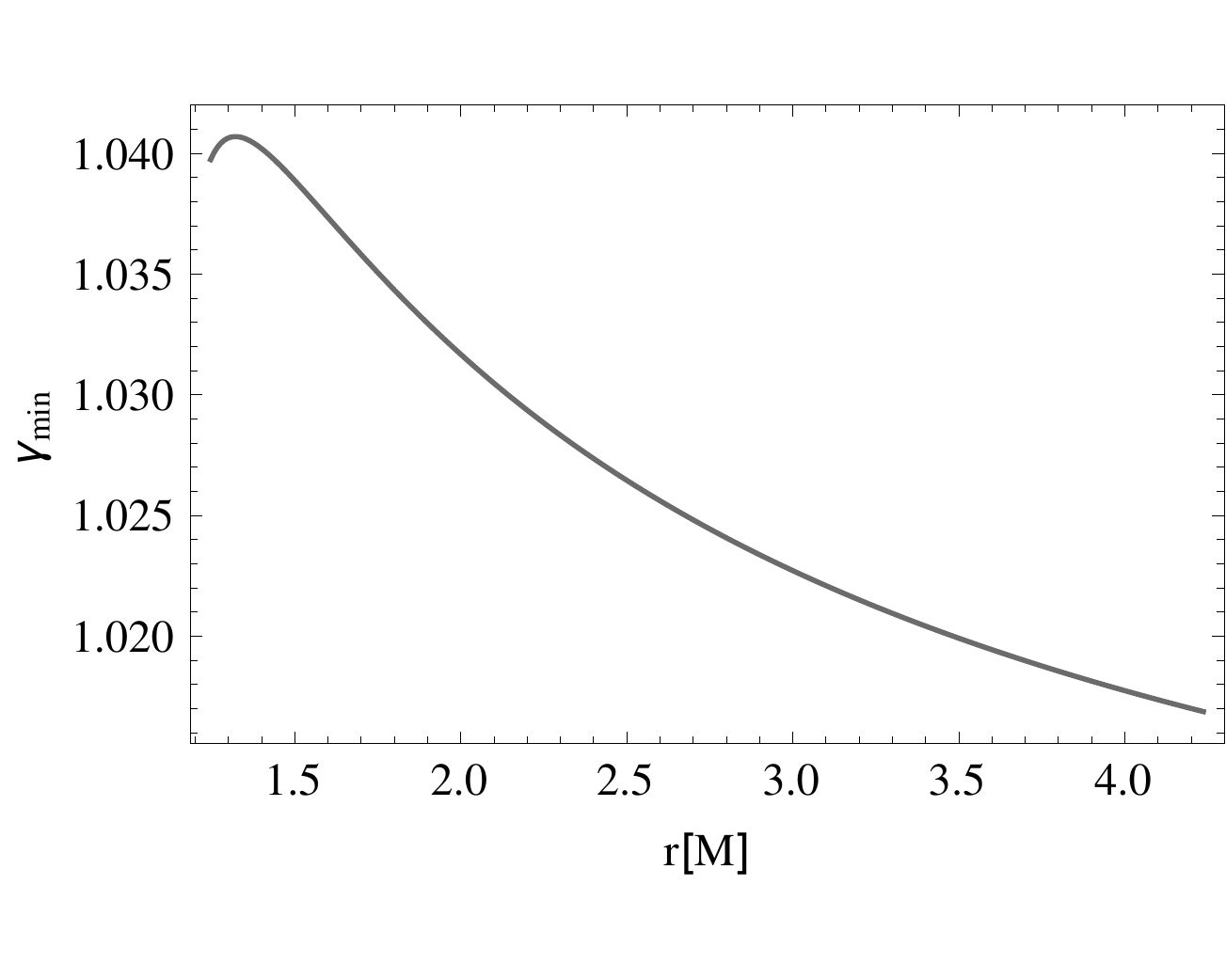}
\caption{Dependence of $\gamma_{min}$ on the radius of an orbit of an orbiting particle for $a=0.998M$.}
\label{fig4}
\end{figure}

\color{black}
\section{Summary}\label{sec5}
A derivation of the lower bound of the Lorentz factor for collisions on orbits near rotating black holes is presented in this paper.
Such a factor can always be minimized to a finite value, even for ISCO orbits for the extreme--Kerr limit. 
In the extremal limit $\gamma_{min}=\frac{2 \sqrt2-1}{\sqrt{3}}\simeq1.056$, what
corresponds to a speed of collision $v_{min}\simeq 0.32 c$. For Thorne bound \footnote{Maximal spin of astrophysical black holes predicted by \cite{Th74}.}
($a=0.998M$) minimal $\gamma$ for ISCO is $\gamma_{min}\simeq1.039$, $v_{min}\simeq0.27c$ and for Schwarzschild black hole ($a=0$)
$\gamma_{min}\simeq1.006$, $v_{min}\simeq0.11c$.
Dependence between $\gamma_{min}$ and $a$ in the full domain of $a$ is presented in Figure \ref{fig3}.

Dependence of $\gamma_{min}$ on $r$, where r is a radius of an orbit, is presented in Figure \ref{fig4}. Interesting fact is that for some $r>r_{isco}$ minimal value of $\gamma$ is greater
than the value of $\gamma_{min}$ for ISCO.

Allowing incoming particle to have nonzero spin leads to a slight change of calculated parameters: for parallel spins $\gamma_{min}$ is lower and for anti--parallel spins $\gamma_{min}$ is higher than
for non-spinning particles. This kind of additional interaction is no--negligible for near--extreme Kerr only.

To conclude, we have shown that in addition to the existence of a lower bound of $\gamma$ in every case, there cannot exist “gentle” collisions with ISCO--orbiting particles and all the collisions occur with a high velocity.
Our result can be used as an initial value for the simulation of head--on neutron stars collisions in order to find out if collapse to the black holes is likely to happen with
such a minimal velocity.
In terms of a planet --- spaceship issue discussed in the introduction, we conclude that it would be impossible to
avoid a huge load factor when landing on an ISCO planet. What is more, energy sufficient to ``slow down'' near the planet would also be beyond present technological capabilites of humankind.
Of course this reasoning does not take into account
the possibility of adjusting a spaceship's trajectory using engines or gravitational assist (i.e. Thorne's idea to reduce the velocity of an impact \cite{Thorne}).

\begin{acknowledgements}
I am grateful to Sebastian Szybka for suggesting the topic and useful comments.
\end{acknowledgements}
\bibliography{article}
\bibliographystyle{spmpsci}      

\end{document}